\documentstyle[aps,prd,epsfig]{revtex}
\begin{document}
\textheight 21.2cm
\setlength{\topmargin}{-1cm}
\pagestyle{plain}

%
 
\def\@maketitle{\newpage
 \null
 \vskip 1em              
 \begin{flushright}
  {\normalsize \@date}           
 \end{flushright}
 \vskip 2em                 
 \begin{center}
  {\Large\bf \@title \par}     
  \vskip 1.5em                
  {\large                        
   \lineskip .5em           
   \begin{tabular}[t]{c}\@author
   \end{tabular}\par}
\end{center}
 \par
 \vskip 1.5em}                
 
\def\abstract{\if@twocolumn
\section*{Abstract}
\else \normalsize
\fi}
 
\def\endabstract{\if@twocolumn\fi\par\clearpage}
 
\parskip 1.5pt plus 1.5pt          
\tabcolsep 12pt                    

%
%
\def\eg{\hbox{\it e.g.}}        \def\cf{\hbox{\it cf.}}
\def\etal{\hbox{\it et al.}}
\def\dash{\hbox{---}}
\def\bR{\mathop{\bf R}}
\def\bC{\mathop{\bf C}}
\def\eq#1{{eq. \ref{#1}}}
\def\eqs#1#2{{eqs. \ref{#1}--\ref{#2}}}
\def\lie{\hbox{\it \$}} 
\def\partder#1#2{{\partial #1\over\partial #2}}
\def\secder#1#2#3{{\partial^2 #1\over\partial #2 \partial #3}}
\def\abs#1{\left| #1\right|}
\def\ltap{\ \raisebox{-.4ex}{\rlap{$\sim$}} \raisebox{.4ex}{$<$}\ }
\def\gtap{\ \raisebox{-.4ex}{\rlap{$\sim$}} \raisebox{.4ex}{$>$}\ }
\def\contract{\makebox[1.2em][c]{
        \mbox{\rule{.6em}{.01truein}\rule{.01truein}{.6em}}}}
%
\def\com#1#2{
        \left[#1, #2\right]}
%
%
\def\bentarrow{\:\raisebox{1.3ex}{\rlap{$\vert$}}\!\rightarrow}
\def\longbent{\:\raisebox{3.5ex}{\rlap{$\vert$}}\raisebox{1.3ex}%
        {\rlap{$\vert$}}\!\rightarrow}
\def\onedk#1#2{
        \begin{equation}
        \begin{array}{l}
         #1 \\
         \bentarrow #2
        \end{array}
        \end{equation}
                }
\def\dk#1#2#3{
        \begin{equation}
        \begin{array}{r c l}
        #1 & \rightarrow & #2 \\
         & & \bentarrow #3
        \end{array}
        \end{equation}
                }
\def\dkp#1#2#3#4{
        \begin{equation}
        \begin{array}{r c l}
        #1 & \rightarrow & #2#3 \\
         & & \phantom{\; #2}\bentarrow #4
        \end{array}
        \end{equation}
                }
\def\bothdk#1#2#3#4#5{
        \begin{equation}
        \begin{array}{r c l}
        #1 & \rightarrow & #2#3 \\
         & & \:\raisebox{1.3ex}{\rlap{$\vert$}}\raisebox{-0.5ex}{$\vert$}%
        \phantom{#2}\!\bentarrow #4 \\
         & & \bentarrow #5
        \end{array}
        \end{equation}
                }
%
%
%
\def\ap#1#2#3{     {\it Ann. Phys. (NY) }{\bf #1} (19#2) #3}
\def\arnps#1#2#3{  {\it Ann. Rev. Nucl. Part. Sci. }{\bf #1} (19#2) #3}
\def\npb#1#2#3{    {\it Nucl. Phys. }{\bf B #1} (19#2) #3}
\def\plb#1#2#3{    {\it Phys. Lett. }{\bf B #1} (19#2) #3}
\def\prd#1#2#3{    {\it Phys. Rev. }{\bf D #1} (19#2) #3}
\def\prep#1#2#3{   {\it Phys. Rep. }{\bf #1} (19#2) #3}
\def\prl#1#2#3{    {\it Phys. Rev. Lett. }{\bf #1} (19#2) #3}
\def\ptp#1#2#3{    {\it Prog. Theor. Phys. }{\bf #1} (19#2) #3}
\def\rmp#1#2#3{    {\it Rev. Mod. Phys. }{\bf #1} (19#2) #3}
\def\zpc#1#2#3{    {\it Zeitschr. f{\"u}r Physik }{\bf C #1} (19#2) #3}
\def\mpla#1#2#3{   {\it Mod. Phys. Lett. }{\bf A #1} (19#2) #3}
\def\sjnp#1#2#3{   {\it Sov. J. Nucl. Phys. }{\bf #1} (19#2) #3}
\def\yf#1#2#3{     {\it Yad. Fiz. }{\bf #1} (19#2) #3}
\def\nc#1#2#3{     {\it Nuovo Cim. }{\bf #1} (19#2) #3}
\def\jetpl#1#2#3{  {\it JETP Lett. }{\bf #1} (19#2) #3}
\def\ib#1#2#3{     {\it ibid. }{\bf #1} (19#2) #3}
\def\lmp#1#2#3{    {\it Lett. Math. Phys. }{\bf #1} (19#2) #3}
\def\app#1#2#3{    {\it Act. Phys. Pol.}{\bf B #1} (19#2) #3}
\def\cqg#1#2#3{    {\it Class. Quant. Grav. }{\bf #1} (19#2) #3}  
\newcommand{\1}{{\'\i}}
\newcommand{\be}{\begin{equation}}
\newcommand{\ee}{\end{equation}\noindent}
\newcommand{\bear}{\begin{eqnarray}}
\newcommand{\ear}{\end{eqnarray}\noindent}
\newcommand{\benn}{\begin{enumerate}}
\newcommand{\enn}{\end{enumerate}}
\newcommand{\no}{\noindent}
\date{}
\renewcommand{\theequation}{\arabic{section}.\arabic{equation}}
\renewcommand{\arraystretch}{2.5}
\newcommand{\GeV}{\mbox{GeV}}
\newcommand{\cL}{\cal L}
\newcommand{\D}{\cal D}
\newcommand{\Dhalf}{{D\over 2}}
\newcommand{\Det}{{\rm Det}}
\newcommand{\PP}{\cal P}
\newcommand{\G}{{\cal G}}
\def\R{1\!\!{\rm R}}
\def\Eins{\mathord{1\hskip -1.5pt
\vrule width .5pt height 7.75pt depth -.2pt \hskip -1.2pt
\vrule width 2.5pt height .3pt depth -.05pt \hskip 1.5pt}}
\newcommand{\symb}{\mbox{symb}}
\renewcommand{\arraystretch}{2.5}
\newcommand{\slD}{\raise.15ex\hbox{$/$}\kern-.57em\hbox{$D$}}
\newcommand{\slpartial}{\raise.15ex\hbox{$/$}\kern-.57em\hbox{$\partial$}}
\newcommand{\slG}{{{\dot G}\!\!\!\! \raise.15ex\hbox {/}}}
\newcommand{\Gd}{{\dot G}}
\newcommand{\Gund}{{\underline{\dot G}}}
\newcommand{\Gdd}{{\ddot G}}
\def\GBd12{{\dot G}_{B12}}
\def\mneg{\!\!\!\!\!\!\!\!\!\!}
\def\Mneg{\!\!\!\!\!\!\!\!\!\!\!\!\!\!\!\!\!\!\!\!}
\def\non{\nonumber}
\def\beqn*{\begin{eqnarray*}}
\def\eqn*{\end{eqnarray*}}
\def\sy{\scriptscriptstyle}
\def\footstrut{\baselineskip 12pt}
\def\square{\kern1pt\vbox{\hrule height 1.2pt\hbox{\vrule width 1.2pt
   \hskip 3pt\vbox{\vskip 6pt}\hskip 3pt\vrule width 0.6pt}
   \hrule height 0.6pt}\kern1pt}
\def\np{n_{+}}
\def\nm{n_{-}}
\def\Np{N_{+}}
\def\Nm{N_{-}}
\def\slash#1{#1\!\!\!\raise.15ex\hbox {/}}
\def\dint#1{\int\!\!\!\!\!\int\limits_{\!\!#1}}
\def\bra#1{\langle #1 |}
\def\ket#1{| #1 \rangle}
\def\vev#1{\langle #1 \rangle}
\def\rightvac{\mid 0\rangle}
\def\leftvac{\langle 0\mid}
\def\dps{\displaystyle}
\def\sy{\scriptscriptstyle}
\def\half{{1\over 2}}
\def\third{{1\over3}}
\def\fourth{{1\over4}}
\def\fifth{{1\over5}}
\def\sixth{{1\over6}}
\def\seventh{{1\over7}}
\def\eigth{{1\over8}}
\def\ninth{{1\over9}}
\def\tenth{{1\over10}}
\def\pa{\partial}
\def\ddtau{{d\over d\tau}}
\def\ge{\hbox{\textfont1=\tame $\gamma_1$}}
\def\gz{\hbox{\textfont1=\tame $\gamma_2$}}
\def\gd{\hbox{\textfont1=\tame $\gamma_3$}}
\def\go{\hbox{\textfont1=\tamt $\gamma_1$}}
\def\gt{\hbox{\textfont1=\tamt $\gamma_2$}}
\def\gth{\hbox{\textfont1=\tamt $\gamma_3$}} 
\def\gf{\hbox{$\gamma_5\;$}}
\def\ie{\hbox{$\textstyle{\int_1}$}}
\def\iz{\hbox{$\textstyle{\int_2}$}}
\def\id{\hbox{$\textstyle{\int_3}$}}
\def\ldop{\hbox{$\lbrace\mskip -4.5mu\mid$}}
\def\rdop{\hbox{$\mid\mskip -4.3mu\rbrace$}}
\def\eps{\epsilon}
\def\epshalf{{\epsilon\over 2}}
\def\e{\mbox{e}}
\def\g{\mbox{g}}
\def\kinb{{1\over 4}\dot x^2}
\def\kinf{{1\over 2}\psi\dot\psi}
\def\expk{{\rm exp}\biggl[\,\sum_{i<j=1}^4 G_{Bij}k_i\cdot k_j\biggr]}
\def\expp{{\rm exp}\biggl[\,\sum_{i<j=1}^4 G_{Bij}p_i\cdot p_j\biggr]}
\def\expshort{{\e}^{\half G_{Bij}k_i\cdot k_j}}
\def\expabb{{\e}^{(\cdot )}}
\def\epseps#1#2{\varepsilon_{#1}\cdot \varepsilon_{#2}}
\def\epsk#1#2{\varepsilon_{#1}\cdot k_{#2}}
\def\kk#1#2{k_{#1}\cdot k_{#2}}
\def\G#1#2{G_{B#1#2}}
\def\Gp#1#2{{\dot G_{B#1#2}}}
\def\GF#1#2{G_{F#1#2}}
\def\Dab{{(x_a-x_b)}}
\def\Dsq{{({(x_a-x_b)}^2)}}
\def\lag{( -\partial^2 + V)}
\def\PITD{{(4\pi T)}^{-{D\over 2}}}
\def\4piTD{{(4\pi T)}^{-{D\over 2}}}
\def\4piT4{{(4\pi T)}^{-2}}
\def\TintmD{{\dps\int_{0}^{\infty}}{dT\over T}\,e^{-m^2T}
    {(4\pi T)}^{-{D\over 2}}}
\def\Tintm4{{\dps\int_{0}^{\infty}}{dT\over T}\,e^{-m^2T}
    {(4\pi T)}^{-2}}
\def\Tintm{{\dps\int_{0}^{\infty}}{dT\over T}\,e^{-m^2T}}
\def\Tint{{\dps\int_{0}^{\infty}}{dT\over T}}
\def\pint{{\dps\int}{dp_i\over {(2\pi)}^d}}
\def\Dx{\dps\int{\cal D}x}
\def\Dy{\dps\int{\cal D}y}
\def\Dpsi{\dps\int{\cal D}\psi}
\def\Tr{{\rm Tr}\,}
\def\tr{{\rm tr}\,}
\def\sumij{\sum_{i<j}}
\def\freeexp{{\rm e}^{-\int_0^Td\tau {1\over 4}\dot x^2}}
\def\arraystretch{2.5}
\def\Ge{\mbox{GeV}}
\def\dA{\partial^2}
\def\DA{\sqsubset\!\!\!\!\sqsupset}
\def\FFdual{F\cdot\tilde F}
\def\mn{{\mu\nu}}
\def\rs{{\rho\sigma}}
%
\def\ap#1#2#3{     {\it Ann. Phys. (NY) }{\bf #1} (19#2) #3}
\def\arnps#1#2#3{  {\it Ann. Rev. Nucl. Part. Sci. }{\bf #1} (19#2) #3}
\def\npb#1#2#3{    {\it Nucl. Phys. }{\bf B #1} (19#2) #3}
\def\plb#1#2#3{    {\it Phys. Lett. }{\bf B #1} (19#2) #3}
\def\prd#1#2#3{    {\it Phys. Rev. }{\bf D #1} (19#2) #3}
\def\prep#1#2#3{   {\it Phys. Rep. }{\bf #1} (19#2) #3}
\def\prl#1#2#3{    {\it Phys. Rev. Lett. }{\bf #1} (19#2) #3}
\def\ptp#1#2#3{    {\it Prog. Theor. Phys. }{\bf #1} (19#2) #3}
\def\rmp#1#2#3{    {\it Rev. Mod. Phys. }{\bf #1} (19#2) #3}
\def\zpc#1#2#3{    {\it Zeitschr. f{\"u}r Physik }{\bf C #1} (19#2) #3}
\def\mpla#1#2#3{   {\it Mod. Phys. Lett. }{\bf A #1} (19#2) #3}
\def\sjnp#1#2#3{   {\it Sov. J. Nucl. Phys. }{\bf #1} (19#2) #3}
\def\yf#1#2#3{     {\it Yad. Fiz. }{\bf #1} (19#2) #3}
\def\nc#1#2#3{     {\it Nuovo Cim. }{\bf #1} (19#2) #3}
\def\jetpl#1#2#3{  {\it JETP Lett. }{\bf #1} (19#2) #3}
\def\ib#1#2#3{     {\it ibid. }{\bf #1} (19#2) #3}
\def\lmp#1#2#3{    {\it Lett. Math. Phys. }{\bf #1} (19#2) #3}
\def\app#1#2#3{    {\it Act. Phys. Pol.}{\bf B #1} (19#2) #3}
\def\cqg#1#2#3{    {\it Class. Quant. Grav. }{\bf #1} (19#2) #3}  
%
%
%
\def\bbbr{{\rm I\!R}}
\def\bbbone{{\mathchoice {\rm 1\mskip-4mu l} {\rm 1\mskip-4mu l}
{\rm 1\mskip-4.5mu l} {\rm 1\mskip-5mu l}}}
\def\bbbz{{\mathchoice {\hbox{$\sf\textstyle Z\kern-0.4em Z$}}
{\hbox{$\sf\textstyle Z\kern-0.4em Z$}}
{\hbox{$\sf\scriptstyle Z\kern-0.3em Z$}}
{\hbox{$\sf\scriptscriptstyle Z\kern-0.2em Z$}}}}
%

\pagestyle{empty}
\renewcommand{\thefootnote}{\fnsymbol{footnote}}
\hskip 13cm {\sl UMSNH-Phys/01-9}   
\vskip .4cm
\begin{center}
{\Large\bf Closed-form two-loop 
Euler-Heisenberg Lagrangian}\\
{\Large\bf in a self-dual background}
\vskip1.3cm
{\large Gerald V. Dunne}
\\[1.5ex]
{\it
Department of Physics\\
University of Connecticut\\
Storrs CT 06269, USA
}
\vspace{.8cm}

 {\large Christian Schubert
}
\\[1.5ex]
{\it
Instituto de F\'{\i}sica y Matem\'aticas
\\
Universidad Michoacana de San Nicol\'as de Hidalgo\\
Apdo. Postal 2-82\\
C.P. 58040, Morelia, Michoac\'an, M\'exico\\
schubert@itzel.ifm.umich.mx\\
}
\vspace{.1cm}
and\\
\vspace{.1cm}
{\it
California Institute for Physics and Astrophysics\\
366 Cambridge Ave., Palo Alto, CA 94306, USA}

\vskip 2cm
 {\large \bf Abstract}
\end{center}
\begin{quotation}
\noindent

We show that the two-loop Euler-Heisenberg effective Lagrangian for scalar
QED in a constant Euclidean self-dual background has a simple explicit
closed form expression in terms of the digamma function. This result leads
to a simple analysis of the weak- and  strong-field expansions, the
two-loop scalar QED beta function, and the analytic continuation
properties of the effective Lagrangian and its imaginary part. 
\end{quotation}
\vskip 1cm
\clearpage
\renewcommand{\thefootnote}{\protect\arabic{footnote}}
\pagestyle{plain}

\setcounter{page}{1}
\setcounter{footnote}{0}
\renewcommand{\theequation}{\arabic{equation}}
\setcounter{equation}{0}

The Euler-Heisenberg Lagrangian, one of the earliest results
in quantum electrodynamics \cite{eulhei,weisskopf}, describes the
complete one-loop amplitude involving a spinor loop
interacting non-perturbatively with a constant background
field. This effective Lagrangian encodes the information concerning the
one-loop amplitude in a form which is extremely convenient for the study
at low energies of nonlinear QED effects such as photon-photon scattering
\cite{eulhei}, photon dispersion \cite{adler71}, and photon splitting
\cite{adler71}. Analogous results apply for scalar QED
\cite{schwinger51}. The Euler-Heisenberg Lagrangian is real for a purely
magnetic field, while in the presence of an electric field there is an
absorptive part, indicative of the possibility of electron-positron pair
creation by the field \cite{schwinger51,nikishov}. Although pair
production is exponentially suppressed for field strengths which are
presently possible for a macroscopic field in the laboratory, pair
creation has recently been observed in an experiment involving electrons
traversing the focus of a terawatt laser \cite{burkeetal}. Moreover, in
the near future the construction of X-ray free electron lasers may allow
one to obtain field strengths much higher than the ones obtainable by
optical lasers \cite{ringwald}.

All these results are at the one-loop level. Going to the two-loop
level, the first radiative correction to the Euler-Heisenberg Lagrangian,
describing the effect of an additional photon exchange in the loop, was
first studied in the seventies by Ritus \cite{ritus1,ritus2,ritus3}, for
both spinor and scalar QED, and for a general constant background field
$F_{\mu\nu}$. Using the exact propagators in a constant field found by
Fock \cite{fock} and Schwinger \cite{schwinger51}, and a proper-time
cutoff as the UV regulator, Ritus obtained the two-loop effective
Lagrangian in terms of a certain two-parameter integral. A
new feature of the two-loop calculation is the necessity of mass
renormalization, which turns out to work in a quite non-trivial
way \cite{ritus1,ritus2,ritus3,ditreu,rss,frss,ds}.
Unfortunately, the double parameter integral is very
complicated, so it is much more difficult to study the weak- and
strong-field expansions at two-loops than at one-loop. This is true even
for the special cases where the background is purely magnetic or purely
electric
\cite{ritus1,ritus2,ritus3}. More recently, the world-line approach to
quantum field theory 
\cite{berkos,strassler,mckshe,ss3,shaisultanov,review} 
has been used
\cite{rss,frss,korsch} 
to re-compute these two-loop results, for both spinor and
scalar QED, in a general constant background field. The world-line results
are once again expressed as complicated double parameter integrals,
although they are of a different form from Ritus's expressions.
While it has not yet been shown how to convert between these two
different expressions, their equivalence has been checked, as weak- and
strong-field expansions, using computer-based expansions. The Borel
summation properties of the weak-field expansion have also been studied
\cite{ds}.

In this letter we specialize the world-line two-loop calculation to a
constant Euclidean self-dual background, where $F_{\mu\nu}=\tilde
F_{\mu\nu}\equiv \half\varepsilon_{\mu\nu\alpha\beta}F^{\alpha\beta}$.
In this case we find the surprising result that {\it both} of the
parameter integrals can be done in closed form, yielding a simple
expression (see Eq.[\ref{sd2}]) for the scalar QED effective Lagrangian,
in terms of the digamma function $\psi(x)=d/dx\,\ln\Gamma(x)$. This
simple closed form makes it extremely easy to study the weak- and
strong-field expansions, as well as analyticity features such as analytic
continuation and Borel summation. We have chosen to consider scalar QED
rather than  spinor QED as it leads to some computational
simplification in the world-line formalism, but analogous
simplifications occur for spinor QED \cite{ds2}.

A constant Euclidean self-dual background is of interest for a number of
reasons. First, such backgrounds are significant in QCD, since constant
gauge fields are only stable under quantum fluctuations if they are
self-dual and essentially abelian ({\it i.e.}, have a
fixed direction in color space) \cite{leutwyler}. Second, a constant
self-dual background is relevant for the derivative expansion for an
instanton background in a non-abelian theory
\cite{kwon}. Third, this type of background plays an interesting role in
the interpretation of D-instantons and D3-branes in ${\cal N}=4$
super-Yang-Mills theory \cite{tseytlin,bupets}. Fourth, interesting properties
of the $SU(2)$ effective potential have been found recently for such self-dual
backgrounds \cite{alex}. Finally, from a computational
perspective, the self-dual case is the simplest nontrivial constant field
background, and we will see that it leads to the most explicit expressions
for the effective Lagrangian; hence we hope that it may lead to progress
concerning the general properties of radiative corrections to the
Euler-Heisenberg Lagrangian. 

We begin by briefly recalling the one-loop results for scalar QED. The
one-loop Euler-Heisenberg effective Lagrangian for scalar QED in a
general constant background field $F_{\mu\nu}$ can be expressed as
\cite{schwinger51}
\begin{eqnarray}
{\cal L}_{\rm scalar}^{(1)} =\frac{1}{(4\pi)^2} \int_0^\infty
\frac{dT}{T}\,\e^{-m^2T} \left[\frac{e^2 a\, b}{\sinh(e a T) \sin(e b T)}
-\frac{1}{T^2}+\frac{e^2(a^2-b^2)}{6}\right]
\label{sc1}
\end{eqnarray}
Here $T$ denotes the (Euclidean) proper-time of the
loop fermion, and $a,b$ are related to the two Lorentz
invariants of the Maxwell field by
$a^2-b^2={\bf B}^2-{\bf E}^2,$ and $ab = {\bf E}\cdot{\bf B}$. 
The last term in the square brackets in (\ref{sc1}) corresponds to the
one-loop charge renormalization \cite{schwinger51}. 

If the field has an electric component then this
effective Lagrangian has an imaginary part,
indicating that the vacuum
becomes unstable in an electric field.
A convenient representation
for this imaginary part was found by Schwinger 
\cite{schwinger51}.
In the case of a purely electric field $E$ it reads
\bear
{\rm Im}{\cal L}_{\rm scalar}^{(1)}(E) 
&=&
-\frac{e^2 E^2}{16\pi^3}
\, \sum_{k=1}^\infty \frac{(-1)^{k}}{k^2}
\,\exp\left[-\frac{m^2\pi k}{eE}\right]
\label{L1scalim}
\ear
In this representation the coefficient of the $k$-th
exponential can be directly identified with the
rate for the production of $k$ coherent
pairs by the field \cite{schwinger51,nikishov}. 

For a constant
Euclidean self-dual (SD) background field one can write 
$F^2 = -f^2\Eins$,
so that $a=f$, and $b=-if$. Then, specializing
(\ref{sc1}) to such a background,

\begin{eqnarray}
{\cal L}_{\rm scalar}^{(1)(SD)}(\kappa)
=\frac{m^4}{(4\pi)^2}\,\frac{1}{4\kappa^2}\,
\int_0^\infty \frac{dt}{t^3}\,\e^{-2\kappa t} \left[\frac{t^2}{\sinh^2(t)}
-1+\frac{t^2}{3}\right]
\label{scsd1}
\end{eqnarray}
where we have defined the dimensionless parameter
\begin{eqnarray}\kappa\equiv \frac{m^2}{2 e \sqrt{f^2}} 
\label{kappa}
\end{eqnarray}
and rescaled $t=T e\sqrt{f^2}$.

The one-loop effective Lagrangian in (\ref{scsd1}) may  be expressed
in terms of special functions (such as zeta functions) in various ways
\cite{matt}. However, in anticipation of a comparison with our
two-loop result below, we present the following representation in terms
of the function $\ln\Gamma(x)$ and its integral:
\bear
{\cal L}_{\rm scalar}^{(1)(SD)}(\kappa) 
&=& {m^4\over (4\pi)^2}\frac{1}{\kappa^2}\left[-{1\over
12}\ln \kappa +\zeta'(-1)+\Xi(\kappa)\right]
\label{sd1xi}
\ear 
Here we have defined the function \cite{barnes}
\bear
\Xi(x)\equiv 
-x\ln\Gamma(x)+{x^2\over 2}\ln x-{x^2\over 4}-{x\over 2}+\int_0^x
dy\,\ln\Gamma(y)
\label{defxi}
\ear
and $\zeta^\prime(-1)\approx -0.16542$. It is straightforward to deduce
the weak- and strong-field expansions of the one-loop effective
Lagrangian ${\cal L}_{\rm scalar}^{(1)(SD)}(\kappa)$. The weak field
({\it i.e.}, large $\kappa$) expansion follows most easily from the
integral representation (\ref{scsd1}):
\begin{eqnarray}
{\cal L}_{\rm scalar}^{(1)(SD)}(\kappa)
=\frac{m^4}{(4\pi)^2}\,\sum_{n=2}^\infty c_n^{(1)}\frac{1}{\kappa^{2n}}
\label{sd1weak}
\end{eqnarray}
where the expansion coefficients are (here ${\cal B}_n$
are Bernoulli numbers)
\begin{eqnarray}
c_n^{(1)}=- \frac{{\cal B}_{2n}}{2n(2n-2)}\quad , \quad n=2, 3, \dots
\label{cn1}
\end{eqnarray}
The weak-field expansion (\ref{sd1weak}) is a divergent alternating
series, since the leading behavior of $c_n^{(1)}$ at large order is
\begin{eqnarray}
c_n^{(1)}\sim  2 \frac{(-1)^{n}}{(2\pi)^{2n}}\,\Gamma(2n-1)\quad ,\quad
n\to\infty
\label{growth1}
\end{eqnarray}
The strong-field ({\it i.e.}, small $\kappa$) expansion of ${\cal L}_{\rm
scalar}^{(1)(SD)}(\kappa)$ follows most easily from the
representation (\ref{sd1xi}):
\begin{eqnarray}
{\cal L}_{\rm
scalar}^{(1)(SD)}(\kappa)=\frac{m^4}{(4\pi)^2}\,\frac{1}{\kappa^2}\,
\left[\left(-\frac{1}{12}+\frac{\kappa^2}{2}\right)\ln
\kappa +\zeta^\prime(-1)+\frac{\kappa}{2}+
(\frac{\gamma}{2}-\frac{1}{4})\kappa^2-
\sum_{n=2}^\infty \frac{(-1)^n \zeta(n)}{(n+1)}\,
\kappa^{n+1}\right]
\label{sd1strong}
\end{eqnarray}
The leading strong-field term is
\begin{eqnarray}
{\cal L}_{\rm scalar}^{(1)(SD)} \sim \frac{1}{12\pi}\, \frac{e^2}{4\pi}\,
f^2\,\ln\left(\frac{2ef}{m^2}\right)
\label{leading1}
\end{eqnarray}
The scalar QED $\beta$-function at one-loop can be read off from the
coefficient of the $f^2 \ln f$ term
\cite{ritus1,ritus2,ritus3,matinyan}. 
In the conventions of \cite{itzzub},
\begin{eqnarray}
\beta_{\rm scalar}(\alpha)=\frac{\alpha^2}{6 \pi}+\dots
\label{beta1}
\end{eqnarray}
where $\alpha=\frac{e^2}{4\pi}$ is the fine structure constant.
This illustrates the connection between the strong field and
high momentum limits of scalar QED \cite{ritus1,ritus2,ritus3,matinyan}.

Turning to the two-loop case, we briefly recall the world-line
results \cite{rss,frss,korsch} for the two-loop Euler-Heisenberg Lagrangian
in a general constant field $F_{\mu\nu}$:
\begin{eqnarray}
{\cal L}^{(2)}_{\rm scalar}
(F)&=&
{(4\pi )}^{-D}
\Bigl(-{e^2\over 2}\Bigr)
\int_0^{\infty}{dT\over T}e^{-m^2T}T^{-{D\over 2}} 
\int_0^{\infty}d\bar T 
\int_0^T d\tau_a
\int_0^T d\tau_b
\nonumber\\
&\phantom{=}&\times
{\rm det}^{-{1\over 2}}
\biggl[{\sin(eFT)\over {eFT}}
\biggr]
{\rm det}^{-{1\over 2}}
\biggl[
\bar T 
-{1\over 2}
{\cal C}_{ab}
\biggr]
\langle
\dot y_a\cdot\dot y_b\rangle
\quad 
\label{Gamma2scal}
\end{eqnarray}
\noindent
Here $T$ and $\bar T$ represent the scalar
and photon proper-times, and $\tau_{a,b}$ the
endpoints of the photon insertion moving around
the scalar loop. 
${\cal C}_{ab}$ and 
$\langle\dot y_a\cdot\dot y_b\rangle$
are given by
\bear
{\cal C}_{ab}&=& 
{\cal G}_B(\tau_a,\tau_a)
-{\cal G}_B(\tau_a,\tau_b)
-{\cal G}_B(\tau_b,\tau_a)
+{\cal G}_B(\tau_b,\tau_b)
\nonumber\\
\langle
\dot y_a\cdot\dot y_b\rangle
&=&
{\rm tr}
\biggl[
\ddot{\cal G}_{Bab}+{1\over 2}
{(\dot {\cal G}_{Baa}-\dot {\cal G}_{Bab})
(\dot {\cal G}_{Bab}-\dot {\cal G}_{Bbb})
\over
{\bar T -{1\over 2}{\cal C}_{ab}}}
\biggr]\; 
\label{defCabWick}
\ear
\noindent
They are expressed in terms of the worldline Green's
function ${\cal G}_B$ (and its first and second derivatives), where
\cite{rss}:
\bear
{\cal G}_{B}(\tau_1,\tau_2) =
{1\over 2{(eF)}^2}\biggl({eF\over{{\rm sin}(eFT)}}
{\rm e}^{-ieFT\dot G_{B12}}
+ieF\dot G_{B12} -{1\over T}\biggr)
\label{calGBetc}
\end{eqnarray}
\noindent
Here $\dot G_{B12} = {\rm sign}(\tau_1-\tau_2) -2(\tau_1-\tau_2)/T$, and 
a `dot' always refers to a derivative with resepect to the
first variable. 
The formula (\ref{calGBetc}) should be
understood as a power series in the field strength tensor $F_{\mu\nu}$.

We now specialize these two-loop formulas to the case of a
self-dual field. The worldline correlator (\ref{calGBetc}) for such a
field simplifies to \cite{vv}:
\bear
{\cal G}_{B12} =
{T\over 2}
\biggl[
{1\over Z^2}-{\cosh(Z\dot G_{B12})\over Z\sinh(Z)}
\biggr]
\Eins + 
i{T\over 2Z^2}\Bigl[{\sinh(Z\dot G_{B12})\over\sinh(Z)}
-\dot G_{B12}\Bigr]{\cal Z}
\label{calGBetcselfdual}
\ear
where ${\cal Z}_{\mu\nu}\equiv eTF_{\mu\nu}$
and $Z\equiv efT$. 
Contrary to the
magnetic or electric field cases, the resulting integrand in
(\ref{Gamma2scal}) is simple enough to allow one to perform, in closed
form, both the integrals over $\bar T$  and over $\tau_1 -\tau_2$.
The result reads, after a partial integration in $T$,
\bear
{\cal L}^{(2)(SD)}_{\rm scalar}(\kappa)= 
{m^4\over(4\pi)^4}\frac{e^2 }{4\kappa^2}
\int_0^{\infty}{dt\over t^3}\,\e^{-2\kappa t}
\Bigl({t\over\sinh(t)}\Bigr)^2
\biggl\lbrace
-4t^2
+6\kappa\, t\Bigl[1-\gamma +\ln\Bigl({1\over 2\kappa \,\sinh(t)}\Bigr)
\Bigr]
\biggr\rbrace
\label{Lfinal}
\ear
Here we have subtracted the appropriate mass renormalization
term, as explained in \cite{frss}.
The (on-shell) renormalization of the Lagrangian is 
completed by further subtracting from the
integrand the terms of order $O(f^0)$,$O(f^2)$. 
The resulting finite 
proper-time integral in (\ref{Lfinal}) can then again
be performed by elementary means, leading to the
following simple expression for the renormalized two-loop effective
Lagrangian:
\bear
{\cal L}_{\rm scalar}^{(2)(SD)}(\kappa)= 
{m^4\alpha\over (4\pi)^3}\,\frac{1}{\kappa^2}\left[
{3\over 2}\xi^2 (\kappa)-\xi'(\kappa)
\right]
\label{sd2}
\ear
where we define [here $\psi(x)$ is the digamma function:
$\psi(x)=\frac{d}{dx}\ln\Gamma(x)$]
\bear
\xi(x)\equiv -x\Bigl(\psi(x)-\ln x +{1\over 2x}\Bigr)
\label{dxi}
\ear
The simplicity of our result (\ref{sd2}) is very surprising, and we
do not know of any comparable two-loop result in gauge theory (for
$\phi^4$ theory an analogous result was obtained in \cite{ilitma}).
Comparing with the one-loop self-dual result (\ref{sd1xi}), we note that
the two-loop result (\ref{sd2}) is very similar, since
$\xi(x)=\Xi^\prime(x)$. In fact, in some sense the two-loop result is even
simpler than the one-loop result, as it involves derivatives of
$\ln\Gamma(x)$ rather than its integral.  

It is now straightforward to derive the weak- and strong-field expansions of
the two-loop effective Lagrangian, using the known expansions of the digamma
function:
\begin{eqnarray}
\psi(x)=\cases{ \ln x-\frac{1}{2x}-\sum_{n=1}^\infty
\frac{{\cal B}_{2n}}{2 n x^{2n}}
\quad , \quad x\to\infty\cr\cr
 -\frac{1}{x}-\gamma+\sum_{n=2}^\infty(-1)^n
\zeta(n) x^{n-1}\quad ,\quad |x|<1} 
\label{psi}
\end{eqnarray}
Therefore, the weak-field ({\it i.e.}, large $\kappa$) expansion is
\bear
{\cal L}^{(2)(SD)}_{\rm scalar}(\kappa)= 
{m^4\over (4\pi)^2}\, \alpha\pi
\sum_{n=2}^{\infty} c^{(2)}_n \frac{1}{\kappa^{2n}}
\label{sd2weak}
\end{eqnarray}
with two-loop expansion coefficients (for $n=2, 3, \dots$):
\begin{eqnarray}
c^{(2)}_n =
{1\over (2\pi)^2}\biggl\lbrace
\frac{2n-3}{2n-2}\,B_{2n-2}
+\frac{3}{2}\sum_{k=1}^{n-1}
{B_{2k}\over 2k}
{B_{2n-2k}\over (2n-2k)}
\biggr\rbrace
\label{cn2}
\ear
At large order in the weak field expansion the leading behavior is
\begin{eqnarray}
c_n^{(2)}\sim  2 \frac{(-1)^{n}}{(2\pi)^{2n}}\,\Gamma(2n-1)\quad ,\quad
n\to\infty
\label{growth2}
\end{eqnarray}
Thus, the leading growth rate (at large order of perturbation theory) of 
the weak-field expansion coefficients is exactly the same at
two-loop as at one-loop, up to an overall factor of $\alpha\pi$. This is
consistent with the results in \cite{ritus1,ritus2,ritus3,ds} for the
purely magnetic and electric cases. We have also checked (to 24 orders)
that the weak-field expansion coefficients in (\ref{cn2}) agree with those
obtained by an expansion of Ritus's double-parameter integral expression
in \cite{ritus2}.

Similarly, from (\ref{sd2},\ref{dxi},\ref{psi}), the strong-field ({\it i.e.},
small $\kappa$) expansion is
\begin{eqnarray}
{\cal L}_{\rm scalar}^{(2)(SD)}(\kappa)&=& \frac{m^4
\alpha}{(4\pi)^3}\,\frac{1}{\kappa^2}\left[ \,\left(\gamma+\ln
\kappa\right)\left\{-1+\frac{3}{2}\kappa+
\frac{3}{2}\kappa^2(\gamma+\ln\kappa)-3\sum_{n=2}^\infty (-1)^n \zeta(n)
\kappa^{n+1}\right\}-\frac{5}{8}+\frac{\pi^2}{3}\kappa\right.\nonumber\\
&&\left.-\sum_{n=2}^\infty (-1)^n
\left(\frac{3}{2}\zeta(n)+(n+1)\zeta(n+1)\right)\kappa^n 
+\frac{3}{2}\sum_{n=4}^\infty\sum_{l=2}^{n-2}(-1)^n \zeta(n-l)\zeta(l)
\kappa^n\right] 
\label{sd2strong}
\end{eqnarray}
The leading strong-field term is
\begin{eqnarray}
{\cal L}_{\rm scalar}^{(2)(SD)} \sim \frac{1}{4\pi^2}
\left(\frac{e^2}{4\pi}\right)^2 f^2\,\ln\left(\frac{2ef}{m^2}\right)
\label{leading2}
\end{eqnarray}
From the coefficient of the $f^2 \ln f$ term we can read off 
the two-loop
contribution to the scalar QED $\beta$-function, so that to two-loop
order:
\begin{eqnarray}
\beta_{\rm
scalar}=\frac{\alpha^2}{6\pi}+ \frac{\alpha^3}{2\pi^2} +\dots
\label{betaexp}
\end{eqnarray}
Once again, this illustrates the connection between strong-field and
short-distance physics \cite{ritus1,ritus2,ritus3,matinyan}.

As a final application of our result, we consider the analytic
continuation of ${\cal L}_{\rm scalar}^{(2)(SD)}(\kappa)$ under
$\kappa\to i\kappa$ (or, $f\to -i f$). We
obtain the following explicit Schwinger-type formula:
\bear
{\rm Im}{\cal L}^{(2)(SD)}_{\rm scalar}(i\kappa) =\alpha\pi\,
{m^4\over (4\pi)^3}\frac{1}{\kappa^2}
\sum_{k=1}^{\infty}
\biggl[2\pi \kappa\, k-1+
3\kappa^2{\rm Re}\left(\tilde{\psi}(i\kappa)\right)
\biggr]
\exp\left[-{2\pi\kappa\, k}\right]
\label{imag2}
\ear
where $\tilde{\psi}(x)\equiv \psi(x)-\ln (x) +{1\over 2x}$ is the same
function that appears in the definition of $\xi(x)$ in Eq.(\ref{dxi}).
Note that the imaginary part (\ref{imag2}) is  non-perturbative in the
field [recall Eq.(\ref{kappa})]. This is true also for the one-loop case:
\bear
{\rm Im}{\cal L}_{\rm scalar}^{(1)(SD)}(i\kappa)
&=&
{m^4\over (4\pi)^3}\frac{1}{\kappa^2}
\sum_{k=1}^{\infty}
\biggl[\frac{2\pi\kappa}{k}+{1\over k^2}\biggr]
\,\exp\left[-{2\pi\kappa\, k}\right]
\label{imag1}
\ear
It is interesting that one finds a
completely explicit expression for the exponential prefactor in
(\ref{imag2}), for any $k$. This should be contrasted with the two-loop
constant electric field case where only a 
two-parameter integral representation 
is available \cite{ritus1,ritus2,ritus3}, and where a complicated
analysis \cite{lebrit} of the analyticity properties of the integrand
led to an analogous representation for the imaginary part. But only the
leading term in the $\kappa$ expansion of the prefactor could be
determined. A Borel analysis \cite{ds} of the high orders of perturbation
theory, in this two-loop electric case, found a numerical estimate for
the next term in this Ritus-Lebedev $\kappa$ expansion of the prefactor
(for $k=1$), but was unable to go further. Yet, in the self-dual
background studied here, we find a simple expression for the {\it entire}
prefactor, for {\it all} $k$. We note that for a given value of
$\kappa$, the two-loop prefactor in (\ref{imag2}) eventually becomes
larger than the corresponding one-loop prefactor, when the `instanton'
index $k$ is large enough. 
If this were found to be true also in the electric case,
one would have to conclude that the one-loop Schwinger formula 
(\ref{L1scalim}) does not always provide a reliable estimate of
the actual multi-pair production rate, for very large numbers of pairs.

We have also tested \cite{ds2} the Borel
dispersion relation approach on the self-dual case two-loop weak-field
expansion (\ref{sd2weak}), and  find perfect agreement with the leading
($k=1$) term in the exact result (\ref{imag2}). Indeed, one can already
see that the leading terms in (\ref{growth1}) and (\ref{growth2}) are
consistent with the leading large $\kappa$ terms in the imaginary parts
(\ref{imag1}) and (\ref{imag2}).  

To conclude, we  reiterate that we find it extremely surprising that such
a simple closed-form can be obtained for the two-loop effective
Lagrangian. It suggests that even higher
loop orders might be accessible in the self-dual case, and that
the function $\xi(x)$ will reappear in them. This might ultimately
lead to progress in the computation of higher beta function
coefficients (only the first three coefficients are presently
known for scalar QED \cite{broadhurst}), as well as
provide partial information on higher order light-by-light scattering
\cite{bern}. Similar simplifications occur in the case of spinor QED
\cite{ds2}.  A question of obvious interest is whether some of this
simplification carries over also to the nonabelian case. Work in this 
direction is in progress.

\vskip15pt
{\bf Acknowledgements:}
We thank D. Broadhurst, H. Gies, M. Reuter and V. I. Ritus for helpful
conversations and correspondence. CS thanks the Institute
for Advanced Study, Princeton, for hospitality.
GD thanks the US Dept. of Energy for
support through grant DE-FG02-92ER40716. 



\end{document}